\documentclass[aps,prb,preprint,amssymb]{revtex4}
\usepackage{amsmath}
\usepackage{amsfonts}
\usepackage{amssymb}

\usepackage{epsfig}

\begin{document}
\renewcommand{\thefootnote}{\fnsymbol{footnote}} 

\title{Henry Eyring: Statistical Mechanics, Significant Structure Theory, and the Inductive-Deductive Method}

\author{Douglas Henderson}

\affiliation{Department of Chemistry and Biochemistry, Brigham Young University, Provo UT 84602, E-mail: doug@chem.byu.edu}

\date{\today}

\begin{abstract}
\noindent 
Henry Eyring was, and still is, a towering figure in science. Some aspects of his life and science, beginning in Mexico and continuing in Arizona, California, Wisconsin, Germany, Princeton, and finally Utah, are reviewed here.  Eyring moved gradually from quantum theory toward statistical mechanics and the theory of liquids, motivated in part by his desire to understand reactions in condensed matter.  Significant structure theory, while not as successful as Eyring thought, is better than his critics realize.  Eyring won many awards.  However, most chemists are surprised, if not shocked, that he was never awarded a Nobel Prize. He joined Lise Meitner, Rosalind Franklin, John Slater, and others, in an even more select group, those who should have received a Nobel Prize but did not.   
\end{abstract}

\maketitle 

\section{Introduction}

The author was pleased and honored to have a part in a symposium honoring Eyring that was part of an American Chemical Society Meeting in Salt Lake City in March, 2009.  This article is based on the author's talk in this symposium.  A portion of this article together with two additional figures will appear in the {\it Bulletin for the History of Chemistry},  hereinafter referred to as the {\it Bulletin} article.  Earlier reminiscences and biographies, including one by Henry Eyring, the scientist, and one by his grandson, Henry J. Eyring, have been published earlier$^{1-6}$. Jan Hayes, the organizer of the symposium and this issue, has pointed out to me that I am a coauthor of the last of Henry's publications.  This is somewhat accidental as my book with him was the second edition of {\it Statistical Mechanics and Dynamics}, the first edition appearing nearly twenty years earlier.  Additionally, the publisher wanted camera ready copy for the second edition and since my employer at the time, IBM, could hardly be expected to have me devote all my time to the preparation of the manuscript for this book, the production of the camera ready copy took five years.  Had the preparation proceeded more quickly, I would not have occupied this position.  

Recently, one of Henry's sons told me that Henry loved me.  This is no great distinction as Henry thought positively of everyone.  However, perhaps he loved some people more than others.  He was a very warm and generous person.  My parents were nervous when they were to meet such an eminent person.  He immediately put them at ease.  In any case, my truthful reply to his son was that I loved him.  Henry treated me as an honorary son.  The two scientists of whom I am most fond, Henry Eyring and John Barker, both treated me as a honorary family member.  For this I am deeply grateful.

\section{Early years}

Henry Eyring was grandson of Henry Eyring and Mary Brommeli, who came to America from Germany and Switzerland, respectively.  His grandparents met as they travelled across the plains to Utah in 1860.  They settled first in St. George in southern Utah and later were sent to northern Mexico to help establish a Mormon settlement.  His grandfather was widely respected for his integrity.  His grandmother spent some time in Berlin where she spent a brief period in jail because she refused to compromise her religious beliefs.  His parents were Edward Eyring and Caroline Romney.  Their son, Henry the scientist, was born in Colonia Juarez.  Grandfather Henry Eyring owned a store.  His father, Edward Eyring was a prosperous rancher with several hundred head of cattle. From this point, when I use the names, Henry Eyring and Henry, I refer to the grandson, the scientist.  

Colonia Juarez is a small town that is located in the Mexican state of Chihuahua and is southwest of El Paso and southeast of the Arizona border crossings. Since most of the readers of this article probably do not know the location of Colonia Juarez, a map of this region of Mexico will appear as Fig. 1 in the forthcoming {\it Bulletin} article and that is also available from the author.  Several Mormon settlements were established in the late nineteenth century.  Only two remain, Colonia Dublan and Colonia Juarez.  Colonia Dublan does not appear on the map as it is now a suburb of Nuevos Casas Grandes, which is a sizable city and easily found in the map.  It is roughly in the center of the map, located on Mexico Highway 10. Colonia Juarez is a small town, in a narrow valley,  southwest of Nuevos Casas Grandes, near the end of short secondary road that was, until recently, a gravel road.

The main industry of the region is fruit orchards.  Today, Colonia Dublan /
Nuevos Casas Grandes is the economic center of the region because there is more flat land and a railroad.  However, Colonia Juarez is the religious/cultural center of the Mormon community and looks like a typical small town in Utah.  The bilingual school, Academia Juarez, that Henry attended will be seen in Fig. 2 of the forthcoming {\it Bulletin}, which is a photograph that I took in 2007 and which is also available from the author.  It is seen in the foreground at the bottom of the hill in this photograph.  The building on the right dates back to Henry's time.  The building on the left is more recent.  Some of the houses of Colonia Juarez are barely visible in the background.  

Henry lived in Colonia Juarez until 1912.  Because of the turmoil of the Mexican Revolution, life became dangerous. The Eyring family and most, if not all, of the Mormons were evacuated by rail to El Paso. The women and children were sent first and the men afterwards.  Henry thought that he should be sent with the men and was disappointed that he was part of the first party. The expectation was that they would return soon.  Some did but the Eyring family decided not to return.  The family settled in Arizona in considerably reduced circumstances.  The family thought that Henry was an American by birth.  It was not until the 1930's that he found that he was not.  Thus, some of his most important and famous work was accomplished while he was a Mexican.  It is reasonable to say that he is probably Mexico's most famous chemist.

It is an interesting aside, for me at least, that Pancho Villa briefly `invaded' the US at Columbus, NM, which is almost due north of Colonia Juarez.  With the reluctant agreement of the Mexican government, General Pershing was sent on an unsuccessful expedition to capture Villa; Colonia Dublan was his headquarters and his guides were some of the local Mormons.   However, Henry had left by then and missed this adventure.  

Henry attended the University of Arizona in 1919.  He obtained a BSc and MSc in mining and metallurgical engineering.  I, too, had a (mercifully) brief career in underground (copper) mining.  Eyring decided that a mining/metallurgy career was not for him and he enrolled as a PhD student in chemistry at the University of California in Berkeley in 1925.  After graduation, he spent two years engaged in teaching and research at the University of Wisconsin in Madison.  There he met and in 1928 married his first wife, Mildred Bennion.  They had three sons, Edward (Ted), Henry B. (Hal), and Harden.  Mildred died in 1969.  During her illness that extended over five years, Henry faithfully cared for her and was with her when she died.  A few years after Mildred's death, he married Winifred Brennan, adopting her youngest daughters. 

\section{Berlin and Princeton}

Following his stay in Wisconsin, in 1929 he was awarded a post doctoral fellowship to work at the Kaiser Wilhelm Institute in Dahlem in southwestern Berlin.  Curiously, my parents lived in Dahlem for a time, as members of the diplomatic corps, while I was a doctoral student of Eyring.  Eyring's original plan was to work with Bodenstein but, perhaps fortunately, Bodenstein was away and he collaborated with Michael Polanyi.  

Quantum mechanics was in its infancy and there was much to be done.  Quantum mechanics had not yet been applied to study reactions.  Eyring and Polanyi$^7$ chose to study the simplest reaction, the replacement reaction, H + H$_2$ $\longrightarrow$ H$_2$ + H by applying the Heitler-London method, including exchange.  This was one of the first applications of quantum mechanics to obtain an energy surface for a reaction and, in my opinion, this was one of his most significant papers.  

After a short period back at Berkeley, he joined the faculty at Princeton, where he stayed until 1946.  There he produced many important results.  He developed his famous reaction rate theory$^8$.  A typical plot of the energy, say as calculated by the method of Eyring and Polanyi, that the reacting molecules must trace is plotted in Fig. 1.  This is Fig. 3 of the forthcoming {\it Bulletin} article.  In this plot, the energy of the reactants is on the left and the energy of the products is on the right.  As the incoming molecule approaches the molecule with which it will react, the energy increases.  This energy barrier must be surmounted, rather like a hiker hiking up to and passing over a pass and then descending.  The energy state of the products is on the right and this energy state may be greater or lesser than or equal to that of the reactants. In Fig. 1, the products have a lower energy; this is irrelevant to our argument.  The height of this barrier is $\Delta E^{\ddagger}$.   As would a hiker, the constituents of the reaction may hesitate briefly at the pass.  Eyring coined the name {\it activated complex} for this chemically unstable species at the top of the barrier.  

It is convenient to use a simpler, perhaps simplistic, nonrigorous derivation than that used by Eyring.  In addition to being simpler, this derivation has the advantage of not requiring that the reaction take place in a dilute gas.  In a canonical ensemble, the probability, $P(E)$, of the system having an energy $E$ is 

\begin{equation}
P(E)=\frac{\exp(-E/RT)}{\int_0^{\infty}\exp(-E/RT)dE},
\end{equation}      
where $R$ is the gas constant and $T$ is the temperature.  The denominator is a normalizing factor that ensures that the total probability is one.  The probability of the system having enough energy to reach the top of the pass is the integral of $P(E)$ from $\Delta E^{\ddagger}$ to infinity.  This gives

\begin{eqnarray}    
P(\Delta E^{\ddagger})=\frac{\int_{\Delta E^{\ddagger}}^{\infty}\exp(-E/RT)dE}{\int_0^{\infty}\exp(-E/RT)dE}  
=\exp(-\Delta E^{\ddagger}/RT).
\end{eqnarray}
This result assumes the canonical system, where the volume and number are constant.  However, in the reaction it is the pressure and chemical potential that are constant.  Hence, it is the Gibbs' free energy, $G$, rather than the energy, that should be used.  We should consider

\begin{equation}
P(\Delta G^{\ddagger})=\exp(-\Delta G^{\ddagger}/RT).
\end{equation}

The mode in the activated complex that takes part in the reaction may be thought of a soft spring.  This mode is soft, with a negative spring constant, because the activated complex is unstable.  Using equipartition of energy for a soft spring, the `frequency' of oscillation, $\nu$, of this spring is given by $h\nu=kT$, where $k$ is Boltzmann's constant, the gas constant per molecule.  Thus, formally the reaction rate constant is the product of $\nu$ and $P(\Delta G^{\ddagger})$, 

\begin{equation}
k'=\frac{kT}{h}\exp(-\Delta G^{\ddagger}/RT).
\end{equation}
Of course, the reactants, on reaching the pass and forming an activated complex, may not cross the pass and form the products.  They may fall back.  Hence, it is often convenient to multiply the exponential in Eq. (3) by a factor, $\kappa$, that is called the transmission coefficient. Although there is no general method of calculating $\kappa$, Eyring's rate theory has been very illuminating and has been used in a wide variety of chemical and biological applications.  Eyring was awarded the National Medal of Science, the Berzelius Medal, the Wolf Prize, and many other prestigious awards for this work but, alas, not a Nobel Prize. 

At Princeton, he started writing his famous book, {\it Quantum Chemistry}$^9$.  This may have been the first book in English that used this title.  The writing took a decade.  Eyring told me that Kimball and Walter never met. In any case, the book became a standard text and was translated into several languages.  It was the book from which I first studied quantum mechanics. Of course, I had encountered quantum mechanics but not as the exclusive subject of a course.  Not only is quantum mechanics covered in this book but it is an excellent reference for special functions and group theory.

\section{Utah}

In 1946, with his wife's encouragement, he accepted the position of Dean of the Graduate School at the University of Utah.  The University of Utah, a long established institution, planned to inaugurate a doctoral program; Henry found the chance to help build this program an irresistible temptation.  In this he was highly successful.  The University of Utah has a very prestigious graduate program.  

Earlier he had developed an interest in the theory of liquids.  This, I assume, resulted  from a desire to extend reaction rate theory from gas phase reactions to reactions in condensed phases.  At the time it was thought that in contrast to gases and solids, there was no satisfactory theory of the liquid state.  It is interesting that this is not true.  The van der Waals theory did provide the basis of a satisfactory theory of liquids but this was not understood until recently.  In any case, until the 1960's the thinking was, since the density of a liquid is not too different from that of a solid, a theory of the solid state would be a promising starting point.  Eyring, and others, developed the cell or lattice theory of liquids.  In reality this is a classical (as opposed to quantum) theory of a solid, due to the higher temperatures of most liquids.  Eyring, and probably others, realized that the entropy of the cell theory lacked a factor of $Nk$.  Eyring coined the term, {\it communal entropy}, and added the missing entropy arbitrarily.  Although arbitrary, this is preferable to ignoring the issue and does give a liquid a different free energy than a solid.   

He went one step further and developed the idea that when a molecule evaporated, it left a hole or vacancy in the quasi-lattice of the liquid.  Thus, for every molecule in the vapor phase, there would be a vacancy in the liquid that mirrored the gas molecule.  If this were literally true the sum of the densities of the liquid and vapor would be a constant, equal to the critical density.  This is not quite correct.  The average density of the two phases is a linear function of the temperature but is not a constant and decreases somewhat as the temperature increases.  Nonetheless, this reasoning provides a simple qualitative explanation of the law of rectilinear diameters. 

He `formalized' his reasoning into the {\it significant structure `theory'}$^{10,11}$ at Utah.  Using the idea that a liquid is a mixture of molecules and vacancies that mimic the vapor molecules, the partition function, $Z$, could be written as 

\begin{equation}
Z=Z_s^{V_s/V}Z_g^{(V-V_s)/V},
\end{equation}
where $Z_s$ and $Z_g$ are the partition functions of the solid and vapor phases, respectively, and $V$ and $V_s$ are the volumes of the liquid and solid phases, respectively.  Eyring used the Einstein theory and ideal gas theory for $Z_s$ and $Z_g$.  The Einstein parameter, $\Theta_E$, and $V_s$ are taken from experiment.  The significant structure theory is a description rather than a theory.  Conventionally, a theory in statistical mechanics relates the properties of a system to the forces between the molecules whereas Eyring's description relates the properties of the liquid to those of the solid and vapor without obtaining either from the intermolecular forces.  This said, Eyring by focussing on the volume on the important variable was on the right track and anticipated later developments, such as perturbation theory of liquids. 

One consequence of Eq. (5) is that the heat capacity, $C$, of monatomic liquid, such as argon  becomes, since, for argon, $T$ greatly exceeds $\Theta_E$, 

\begin{equation}
\frac{C}{Nk}=3\frac{V_s}{V}+\frac{3}{2}\frac{V-V_s}{V}.						
\end{equation}
As is seen in Fig. 2 (Fig. 4 of the forthcoming {\it Bulletin} article), Eq. (6) gives a reasonably good description of the heat capacity.  The heat capacity is a second derivative of the free energy and is difficult to obtain accurately in a theory.  The experimental heat capacity becomes infinite at the critical point.  Equation (6) does not predict this.  Much has been made of this failure.  However, it should be kept in mind that no simple theory predicts the singularity of the heat capacity at the critical point. Some are less successful than Eq. (6).   For example, the augmented van der Waals theory (a widely accepted theory) gives the prediction $C=3Nk/2$.   Later Eyring grafted the renormalization group approach onto Eq. (5) to obtain the singularity.  However, I find this artificial.  

I collaborated with him in his study of liquids by applying the significant structure to liquid hydrogen. I also assisted in the writing of the book, {\it Statistical Mechanics and Dynamics} by Eyring, myself, Betsy Stover, and Ted Eyring.  This book was an outgrowth of the lecture notes prepared by one his first students at Utah, Marilyn Alder.  These notes were mimeographed and bound with a yellow cover and was referred to by students as the {\it yellow peril}.  The book was rather unique in that the first chapter covered the field in an informal way and then the material was repeated more formally in the subsequent chapters.  Needless to say, significant structure theory was included in one of the chapters.  This book was moderately successful.  With Jost, he and I collaborated on a multi-volume treatise on physical chemistry.
   
During his final years, he became interested in cancer both because of Mildred's illness and because of the cancer that ultimately took his life.  Betsy Stover came to him with the observation that the mortality curves of the experimental animals that had been exposed to radiation that caused them to die of bone cancer were striking similar to a Fermi-Dirac distribution.   This suggested to Eyring that this was similar to saturation in adsorption and the rate of mutation that was responsible for the cancer was proportional to the product of the fraction of normal cells multiplied by the fraction of mutated cells.  He and Stover wrote several papers under the general title of the {\it Dynamics of Life} that were based on this idea.

\section{Summary}

As I have mentioned Henry had a warm personality.  At times, he became annoyed with someone (including me) but he never held a grudge.  Also, despite his accomplishments, he never felt he was better than someone else.  I found him to be very kind.  

He was quite athletic.  In his youth he could run very fast.  He tells the story of how he outran some students at the University of Arizona who wished to catch him because of the infraction of a foolish rule.  He continued running throughout his life and raced his students.  He could make a standing jump onto to the top of his desk.  I know of only one other person who could do this.  Roberto Benigni does this in the movie, {\it Life is Beautiful}.  I recall that one day in the summer of 1976, while he and I were collaborating on the second edition of {\it Statistical Mechanics and Dynamics} that he had business in the center of city.  Even though he was in his mid seventies, he walked from the univerity to downtown and back, a distance of about four miles that involved walking up a fairly steep hill on the return journey.   

Many people have conjectured about why he never won a Nobel Prize. Henry J. Eyring in reference 5 wonders if it was because he left Princeton for Utah. This is possible. Certainly, his cheering section of prominent people would have been greater if he had stayed at Princeton.  However, one person at the University of Utah has won a Nobel Prize so it is not impossible to win a Nobel Prize at a `provincial' university.  Others have wondered if the fact that Henry was religious played a role.  Perhaps, it was due to Henry's intuitive style of research that was more fashionable in the 1930's than later.  Peter Debye called Henry's style, {\it the inductive-deductive method}.  Henry's description was that his method of finding the path through the forest was first to cut down all the trees in the forest.  My feeling is that his not being awarded a Nobel Prize is part of the uncertainties of life.  He won many prizes.  He would not have won them if the above considerations were a factor.  The Nobel Prize receives too much attention because of the amount of money involved.  The other prizes are equally important.  In any case, he was beloved by all who knew him.

At Henry's funeral, Neal Maxwell, a friend and neighbor, former university colleague, and church leader, said that Henry taught us how to live well and how to die well.  Not a bad epitaph.

\clearpage

\section*{Figure captions}
\label{sec:caption}
\begin{description}

\item[Figure 1.]
The energy of a reaction along the reaction path.  The zero point energy is included.  The reactants are on the left and the products are on the right and are separated by an energy barrier of height $\Delta E^{\ddagger}$.  At the pass over which the reactants must pass for the reaction to proceed, the reacting molecules form a transient complex that Eyring called an activated complex.

\item[Figure 2.]
Heat capacity of argon, as obtained from Eq. (6), compared with experiment.

 \end{description}

\clearpage

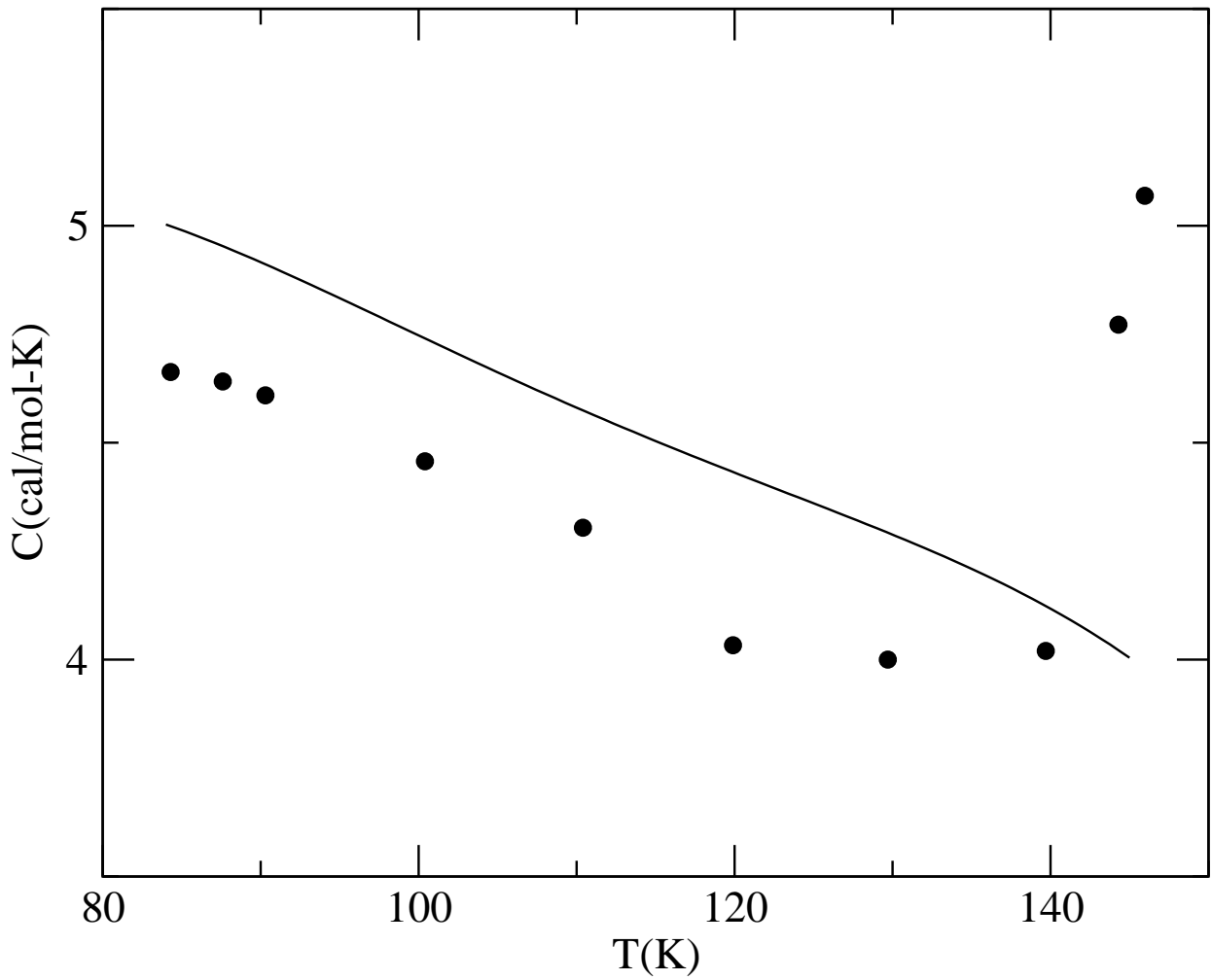
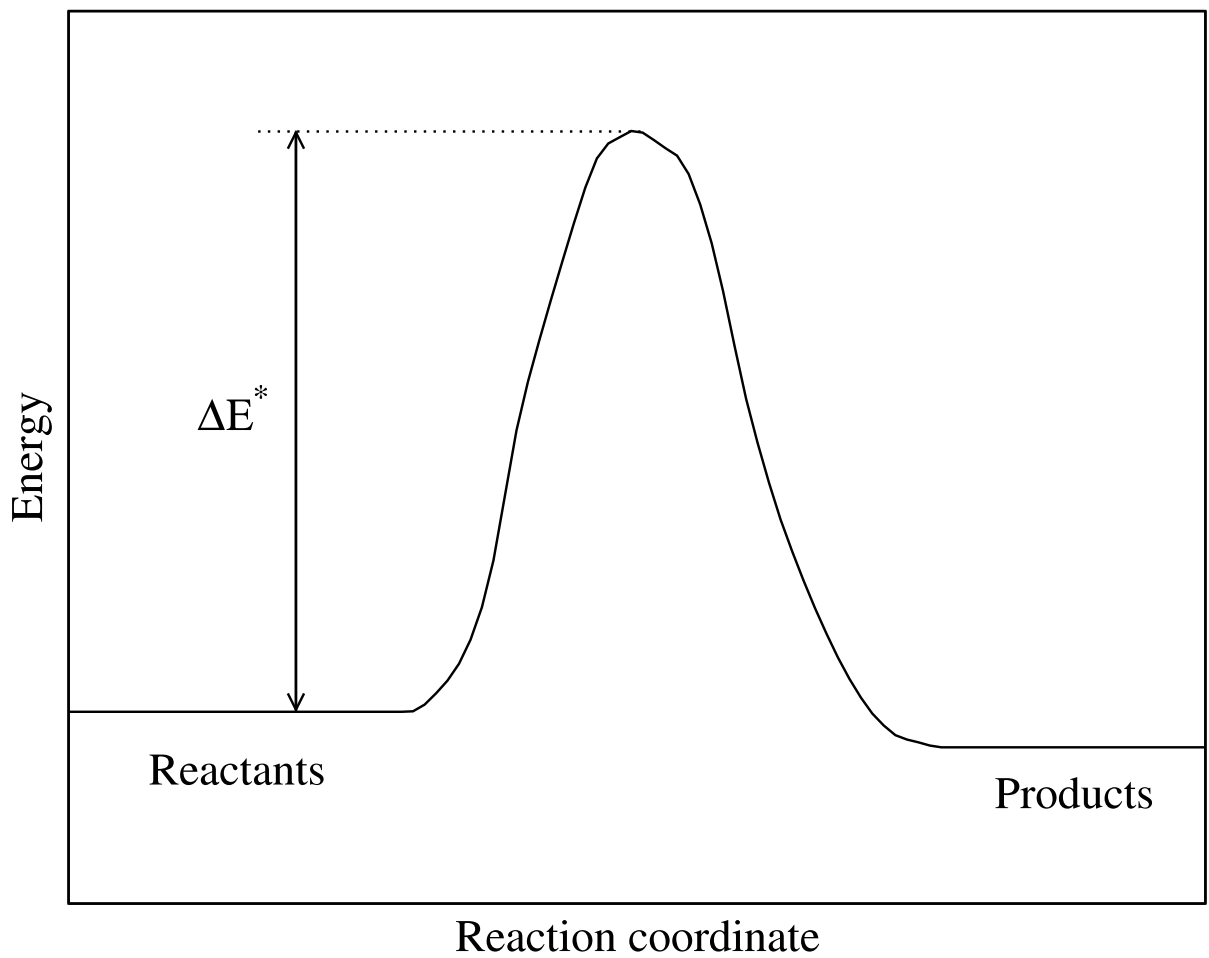
\end{document}